\documentclass[prl,twocolumn]{revtex4}
\usepackage{graphicx}
\def \cH{{\cal{H}} }

\def \nn{\nonumber}

\def \yd{{^\dagger}}
\def \bk{{\bf k}}
\def \bQ{{\bf Q}}

\def \bq{{\bf q}}
\def \bp{{\bf p}}
\def \br{{\bf r}}

\def \up{{\uparrow}}
\def \down{{\downarrow}}
\def \av#1{{\langle#1\rangle}}
\def \momint#1{{\int \frac{d^d #1}{(2\pi)^d}}}

\begin{document}
\title{Enhancement of superconductivity by local inhomogeneity}
\author{Ivar Martin,$^1$ Daniel Podolsky,$^2$ and Steven A. Kivelson$^3$}
\affiliation{$^1$Theoretical Division, Los Alamos National Laboratory, Los
Alamos, NM 87545\\
$^2$Department of Physics, University of California, Berkeley, CA 94720\\
$^3$Department of Physics, Stanford University, Stanford, CA
94305}
\date{Printed \today}

\begin{abstract}
We study the effect of inhomogeneity of the paring interaction or the
background potential on the superconducting transition temperature, $T_c$.  In
the weak coupling BCS regime, we find that inhomogeneity which is
incommensurate with the Fermi surface nesting vectors {\em enhances} $T_c$
relative to its value for the uniform system. For a fixed modulation strength
we find that the highest $T_c$ is reached when the characteristic modulation
length scale is of the order of the superconducting coherence length.
\end{abstract}
\maketitle

Many strongly correlated superconductors, and in particular high-temperature
superconducting (HTSC) cuprates, exhibit inhomogeneous electronic and/or
structural phases at the nanoscale \cite{davis,kapit,yazd}.  The coexistence of
HTSC and inhomogeneity suggests that the underlying inhomogeneities could be at
least partially responsible for the high value of the superconducting
transition temperature.  Emery and one of us proposed that HTSC is related to
frustrated electronic phase separation, commonly expected in strongly
correlated systems \cite{kiv1}.  These ideas for inhomogeneous
superconductivity have been further developed in the context of stripes
\cite{erica,kiv2}.  It is important to distinguish these and related scenarios
for superconductivity creation or enhancement by inhomogeneity from the
conventional weak-coupling coexistence of superconductivity and various density
waves \cite{machida,sach,ivar,ddw}.  In the latter case, the density wave order
inevitably suppresses superconductivity due to the competition for the Fermi
surface electrons.

It is therefore important to understand the nature of the interplay between
superconductivity and inhomogeneities.  A complete description of the interplay
is clearly impossible. However, in the case in which the characteristic energy
scale responsible for the formation of the inhomogeneity is much larger than
the superconducting energy scale (the gap $\Delta$), and where the residual
interactions are  weak, a description based on BCS theory should be reliable.
The purpose of this work is to study the effect of such imposed inhomogeneity
on superconductivity within the BCS framework. The origin of the inhomogeneity
could be either electronic, as in the frustrated phase separation scenario, or
structural, that is caused by local lattice distortions or non-uniform carrier
concentration due to doping irregularities. We will assume that these
structures do not cause Fermi surface nesting either due to the lack of
periodicity (e.g. random doping profile) or due to the periodicity being
incommensurate with the nested momentum transfers (e.g. frustrated phase
separation, or stripes). Under these conditions we generically find that
inhomogeneity $enhances$ the global superconducting transition temperature,
$T_c$.  At the mean-field level, the maximum $T_c$ is achieved when the
characteristic length-scale of the inhomogeneities, $L$,  is large, in which
case the transition temperature is that of the regions with the highest local
$T_c$. Upon including the effects of phase fluctuations we find that $T_c$ is
maximized when $L$ is comparable to the superconducting coherence length $\xi
\sim v_F/T_c$. The increase of the transition temperature occurs at the expense
of the superfluid density, which is reduced in inhomogeneous superconductors
relative to their homogeneous counterparts.

{\em Inhomogeneous pairing: Mean-Field treatment.}  As a first example we
consider a Hubbard model with an inhomogeneous attractive potential $U(\br)>0$,
\begin{eqnarray}
\cH&=&\cH_0+\cH_U
\label{ham} \\
\cH_0&=&\sum_{\bk\sigma} \xi_\bk c\yd_{\bk\sigma}c_{\bk\sigma}\nn\\
\cH_U&=&-\sum_{\br}U(\br)n_\up(\br) n_\down(\br)\nn,
\end{eqnarray}
where $\xi_\bk=\epsilon_\bk-\mu$ and
$n_\sigma(\br)=c\yd_\sigma(\br)c_\sigma(\br)$ is the occupation number of
electrons of spin $\sigma$ at position $\br$.  Within this model, our goal is
to understand whether for a {\em fixed average} pairing strength
$\overline{U(\br)}$, a uniform or non-uniform $U(\br)$ yields a higher
transition temperature, $T_c$.  In the weak coupling limit, we can derive the
BCS condition for the onset of superconductivity from the Hamiltonian
(\ref{ham}),
\begin{eqnarray}
\Delta_\bq=\momint{p}U(\bq-\bp)K(\bp)\Delta_\bp, \label{bcs-cond}
\end{eqnarray}
where $\Delta_\bq=\sum_{\bk\,\bp} U(\bk)
\av{c_{\bq/2-\bk/2+\bp\up}c_{\bq/2-\bk/2-\bp\down}}$, $U(\bk)$ is
the Fourier transform of the pairing interaction, and $K(\bp)$ is
the pairing kernel.  The pairing kernel depends on temperature $T$
and the mean-field (MF) superconducting transition is defined by
the temperature at which the integral equation has a non-trivial
solution. The kernel can be calculated from the normal state
electron Green functions \cite{agd}
\begin{eqnarray}\label{eq:Kq}
K(\bp)\approx N_f\ln\left[\frac{2 \gamma\omega_D}{\pi\sqrt{T^2+(v_f
p)^2}}\right]\Theta(\omega_D-|v_f p|),
\end{eqnarray}
where $N_f$ is the density of states at the Fermi surface, $v_f$
is the Fermi velocity, $T$ is temperature, and $\ln\gamma \approx
0.577$ is Euler's constant. Here we also introduced an explicit
high-energy cut-off for the attraction, $\omega_D$.  For $T\gg v_f
p$ this expression reduces to the well known homogeneous result,
$K\sim N_f\ln [2\gamma\omega_D/(\pi T)]$.

The modulation of the pairing interaction leads to the mixing between Cooper
pairs with different center-of-mass momenta.  For simplicity we first assume a
harmonic modulation ($Q\equiv 2\pi/L$) of the pairing, $U(\br) = {\bar U} +
{U_Q} \cos(\bQ\cdot \br)$. In this case the integral equation (\ref{bcs-cond})
reduces to a system of linear equations $\Delta_n  =  {\bar U}K_n\Delta_n +
({U_Q}/2)[K_{n-1}\Delta_{n-1}+ K_{n+1}\Delta_{n+1}]\equiv {\cal
M}_{nm}\Delta_m$, where $\Delta_n \equiv \Delta(n\bQ + q_0)$ and $K_n \equiv
K(n\bQ + q_0)$.  The ``parent" momentum $q_0$ defines the minimal momentum of a
Cooper pair in the connected family $\Delta(n\bQ + q_0)$.

The paring instability occurs at the temperature $T_c$, such that the largest
eigenvalue of matrix $\cal M$ is equal to 1.  In the uniform case, ${U_Q} = 0$,
this condition is  ${\bar U}K(0) = 1$.  We will now prove that the mean-field
transition temperature is greater than in the uniform case, ${U_Q}=0$. Consider
the $q_0 = 0$ family and  without loss of generality, take ${U_Q}> 0$. Since
all of the matrix elements of $\cal M$ are non-negative, by Perron's theorem
\cite{Bellman}, the maximal eigenvalue is a positive number that is larger than
any diagonal matrix element, including ${\bar U}K(0)$. Thus, generically, the
superconducting onset temperature $T_c$ is increased whenever ${U_Q}\ne 0$.

This result can be understood from an analogy with a quantum mechanical
particle in a tight-binding chain.  Defining a new variable
$\Lambda_n={\bar U}K_n\Delta_n$, the BCS condition takes a simple symmetric form, $
\Lambda_n= ({1}/{{\bar U} K_n})\Lambda_n-({U_Q}/2{\bar U})
\left(\Lambda_{n+1}+\Lambda_{n-1}\right)$.  The ``hopping'' term delocalizes
the particle and thus reduces the ``energy" below its minimal on-site value
$({1}/{{\bar U} K_0})$.  Clearly, this leads to a relative increase of $T_c$.

{\em Large $\bQ$ limit.}  In this limit, the quickly oscillating coupling is
ineffective at mixing different modes, so that the off-diagonal terms in ${\cal
M}$ are rapidly decaying with $n$.  We are then justified in keeping only a
small portion of the matrix surrounding the $n=0$ term. The lowest order
correction to the homogeneous result is obtained by considering couplings
between $\Delta_0$ and $\Delta_{\pm 1}$. The largest eigenvalue in this case is
\begin{eqnarray}
\lambda_{\rm max}&=&\frac{{\bar U}}{2}\left[K_0+K_1+\sqrt{(K_0-K_1)^2+\frac{2K_0K_1
{U_Q}^2}{{\bar U}^2}}\right],\nn
\end{eqnarray}
Given the separation of energy scales, \mbox{$T_c\ll v_fQ\ll\omega_D$}, we
obtain $T_c$ by solving $\lambda_{\rm max}=1$,
\begin{eqnarray}\label{eq:Tc_fast}
T_{c}=\frac{2\gamma}{\pi}\omega_D\exp[-1/N_f({\bar U}+\eta)],
\end{eqnarray}
where $\eta={{U_Q}^2K_1}/[2(1-{\bar U}K_1)]$. While $\eta$ is positive, and
since $K_1 \sim \log[\omega_D/v_fQ]$ decreases with increasing $Q$, so does
$\eta$. For $v_FQ>\omega_D$, Cooper pairs can no longer scatter off the quickly
oscillating coupling landscape, and we recover the critical temperature for the
homogeneous case.

{\em Small $\bQ$ limit.}  In the limit $Q \xi \ll 1$, the global MF transition
temperature is determined by the regions with the strongest pairing
interaction,
\mbox{$T_{c}\approx({2\gamma}/{\pi})\omega_D\exp[-1/N_f({\bar U}+|{U_Q}|)]$}. The
deviations from this result due to finite $Q$ and the effects of the phase
fluctuations are discussed below.

\begin{figure}
\includegraphics[width=3 in,height=2.2in]{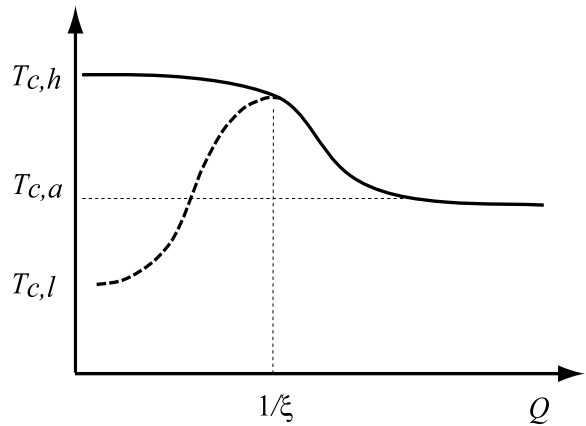}
\caption{Critical temperature for the inhomogeneous negative $U$ Hubbard model
with coupling $U(x) = {\bar U} + {U_Q}\cos(Qx)$.  The thick line denotes the
mean-field result, where $T_{c,a}=(2\gamma/\pi)\omega_D\exp[{-1/{N_f {\bar
U}}}]$ and $T_{c,h}=(2\gamma/\pi)\omega_D\exp[-1/{N_f ({\bar U}+|{U_Q}|)}]$.
The dashed line shows the critical temperature once phase fluctuations of the
order parameter are included. For $Q\xi\ll 1$, the superconductivity is first
established locally in regions where $U(x)$ is large, but macroscopic phase
coherence is achieved at a lower temperature, bounded from below by $T_{c,l} =
(2\gamma/\pi)\omega_D\exp[{-{1}/{N_f ({\bar U}-|{U_Q}|)}}]$.
\label{fig:Tc}\vspace{-3 mm} }
\end{figure}

{\em Electron density modulation.} Before going further, let us in parallel
consider the case of homogeneous coupling $U(\br) = {\bar U}$, with
inhomogeneity caused by a background potential variation.  In the simplest case
of the harmonic modulation, the additional contribution to the Hamiltonian
(\ref{ham}) is
\begin{eqnarray}
\cH_\rho= \rho \sum_{i\sigma} c\yd_{i\sigma} c_{i\sigma}\, \cos \bQ\cdot {\bf
r}_i \label{CDW:Q}.
\end{eqnarray}
It is easy to see that for particle-hole symmetric density of states (DOS), the
linear in $\rho$ contributions to the BCS instability condition equations
vanish identically.  For small values of $\bQ$, the modulation acts as a slowly
varying shift in the local chemical potential with the amplitude proportional
to $|\rho|$. Thus, we only get a linear in $\rho$ contribution for {\em
asymmetric} DOS, $N(\epsilon) = N_f + N^{\prime}(\epsilon - \epsilon_F)$. We then find
that BCS equations are identical to the case of inhomogeneous pairing
interaction with the modulation strength
\begin{eqnarray}
{U_Q}^{eff}=-{\bar U}\frac{N^{\prime}\rho}{N_f}.
\end{eqnarray}

{\em Ginzburg-Landau analysis ($Q\xi \ll 1)$.}  We now consider the general
case of slow variation of the pairing strength and/or background potential. The
Ginzburg-Landau free energy functional in the presence of inhomogeneity is
\begin{eqnarray}\label{eq:F}
F=&&-\int{d{\bf r}\,d{\bf r}^{\prime} K({\bf r}-{\bf r}^{\prime})\Delta({\bf
r})\Delta({\bf r}^{\prime})} + \int{d{\bf r} \frac{\Delta({\bf r})^2}{U({\bf
r})}}\nn\\
&& + \alpha\int{d{\bf r} \rho({\bf r})\Delta({\bf r})^2} +
\frac\beta{2}\int{d{\bf r}\Delta({\bf r})^4},
\end{eqnarray}
Here we assumed that the order parameter remains real even in the presence of
inhomogeneity.  We include both the coupling of the superconducting order
parameter to a density wave, as well as the inhomogeneity of the pairing
interaction.  For small amplitude modulation of the pairing interaction,
$U(\br) = {\bar U} + {\delta U}(\br)$ with $|{\delta U}(\br)| < {\bar U}$, the
two mechanisms are formally equivalent. For particle-hole asymmetric DOS, from
the above considerations, $\alpha = -{\bar U} N^{\prime}/{N_f}$.  For
simplicity, we only consider the inhomogeneous $U(\br)$ case here. The pair
susceptibility kernel is given by Eq.~(\ref{eq:Kq}). In the long wave length
limit,
\begin{equation}
K({\bf r} - {\bf r}^{\prime}) = \delta({\bf r} - {\bf r}^{\prime})
N_f\left[\ln\frac{2\gamma\omega_D}{\pi T} + \xi^2\nabla^2\right],
\end{equation}
where $\xi = v_F/T$. Computing the variation of Eq.~(\ref{eq:F}) with respect
to the order parameter, we find the equation for a stationary solution
$\Delta({\bf r})$,
\begin{eqnarray} \label{eq:GP}
&& -\xi^2\nabla^2\Delta(\br) + g(\br)\Delta(\br) +
\frac{\beta}{N_f}\Delta(\br)^3
= 0,\\
&&\quad\quad\quad g(\br) = \frac{1}{N_f U({\bf r})} -
\ln\frac{2\gamma\omega_D}{\pi T}\nn.
\end{eqnarray}

As a first step, we determine the inhomogeneous mean-field  (MF) transition
temperature for the pairing interaction $U(\br) =  {\bar U} + {U_Q}
\cos(\bQ\cdot\br)$, with $Q\xi \ll 1$.  Close to the MF transition, the cubic
terms in Eq.~(\ref{eq:GP}) can be neglected. Expanding in ${U_Q}/{\bar U}$, and
transforming Eq.~(\ref{eq:GP}) to Fourier space, we obtain a system of
equations connecting $\Delta(\bk)$ and $\Delta(\bk\pm\bQ)$.   For small $Q$,
$\Delta(\bk\pm\bQ) \approx \Delta(\bk)$, and after expanding up to the second
order in $Q$, we obtain
\begin{equation}
-g_{max}\Delta(\bk) =(\xi \bk)^2\Delta(\bk) - \frac12 A^2 Q^2
\partial_k^2\Delta(\bk).
\end{equation}
Here $g_{max}$ denotes $g(\br)$ evaluated at $U(\br) = {\bar U} + |U_Q|$ and $A
= \sqrt{{U_Q}/(N_f {\bar U}^2)}$ (note that there is no explicit constraint on
the value of parameter $A$ since it is a ratio of two small numbers).  The MF
transition temperature is determined by the smallest eigenvalue of the
differential operator on the rhs. This eigenvalue corresponds to the ``ground
state energy'' of a harmonic oscillator, $\xi QA/\sqrt{2}$. The corresponding
transition temperature $T_c^{MF} = T_{max}^{MF}\exp\left(-\xi
QA/\sqrt{2}\right)$, is only slightly less than the transition temperature,
$T^{MF}_{max}$ for a system with a homogeneous pairing interaction
$U_{max}={\bar U}+|{U_Q}|$.  More importantly, it is easy to see that in the
limit of small $Q\xi$, the order parameter is exponentially suppressed in the
region of smaller pairing interaction relative to its value at the peak,
\begin{equation}
\Delta_{\rm min} \sim \Delta_{\rm max} \exp\left\{-A\ (\xi Q)^{-1}\right\},
\label{min}
\end{equation}
This, in turn, implies that phase fluctuations, which we discuss below, can
reduce the global phase coherence temperature significantly below $T_c^{MF}$.

The expression of Eq.~(\ref{min}) is only valid at the MF transition
temperature, where $\Delta$ is infinitesimal.  To determine $\Delta$ below
$T_c^{MF}$ we need to solve the non-linear Eq.~(\ref{eq:GP}).  In a
$d$-dimensional superconductor, with arbitrary smooth variation of $U(\br)$,
the boundary of the ``classically forbidden" region, $g(\br)> 0$,  is a
$(d-1)$-dimensional surface. Hence, near the boundary the problem is
essentially one-dimensional, and in the $g(\br)> 0$ region we can apply the
standard WKB approximation to solve the linearized  Eq.~(\ref{eq:GP}). The
prefactor is fixed by matching the WKB solution to the intermediate asymptotic
at the boundary $x_0$, which can be obtained by solving the full
Eq.~(\ref{eq:GP}) in a linear potential $g(x) = g^\prime(x_0) (x- x_0)$. We
then find that in the particular case of harmonic modulation discussed above,
the order parameter distance $d$ away from the boundary is approximately
\begin{equation}
\Delta(d) \sim T \sqrt{A Q\xi}\exp\left[-A\sqrt{Q\xi}\, (d/ \xi)^{3/2}\right].
\label{eq:min2}
\end{equation}
This expression is obtained assuming  $g^\prime(x_0) \approx A^2 Q$, and
therefore valid only for the temperatures sufficiently below $T_c^{MF}\sim
T_{max}^{MF}$.  So long as $T \gg T_{min}^{MF}$ ($T_{min}^{MF}$ is the uniform
$T_c$ of a system with pairing strength $\bar U - |U_Q|$), the distance from
the turning point to the minimum point of $\Delta$ is $d\sim L$.   Notice, that
this expression depends on temperature not only explicitly, but also through
\mbox{$\xi = v_f/T$}.

{\em Phase fluctuation effects (still with $Q\xi \ll 1$)}. A consequence of the
large spatial variations  in the mean-field $\Delta(\br)$ is that fluctuation
effects are severe where $\Delta(\br)$ is small.  Of these, the most important
fluctuations are thermal fluctuations in the phase of the order parameter, {\it
i.e.} where $\Delta(\br) = |\Delta_{MF}(\br)| e^{i\theta}$ where
$\Delta_{MF}(\br)$ is the solution of Eq. (\ref{eq:GP}), and $\theta(\br)$ is
assumed to be a slowly varying function of $\br$.   The free energy cost of
such phase fluctuations can then be readily computed
\begin{equation}
F_\theta= \int d{\br} J(\br)(\nabla \theta)^2
\end{equation}
where the local superfluid stiffness is \mbox{$J(\br)=
N_f\xi^2|\Delta_{MF}(\br)|^2$}. In general, the phase ordering temperature
estimated using this  as the effective Hamiltonian is reduced from the
mean-field transition temperature ({\it i.e.} the temperature at which $J(\br)$
vanishes), but by an amount that depends on dimensionality, and on the spatial
arrangement of the regions of suppressed stiffness.

For concreteness, we consider the case of a two-dimensional superconductor.  At
finite temperature, no true long-range order is possible \cite{MW}. However, at
$T <T_{KT}$, binding of topological excitations into vortex-antivortex pairs
leads to a state with quasi-long range order, which has a non-zero superfluid
stiffness\cite{KT}. While for homogeneous BCS superconductors in 2D, the
difference between MF and the Kosterlitz-Thouless (KT) transition temperatures
is tiny, \mbox{$(T_c^{MF}-T_{KT})/T_c^{MF}\sim T_c^{MF}/T_F$} (where $T_F$ is
the Fermi temperature), for inhomogeneous superconductors, the suppression of
$T_{KT}$, is generally much larger.  For a smooth random  distribution of
$J(\br)$, an estimate of $T_{KT}$ can be made based on the effective superfluid
density,
\begin{eqnarray}\label{eq:T_KT}
T_{KT} &&\sim \sqrt{{\overline{J(\br)}}\ {[\overline{1/J(\br)}]}^{-1}} \approx
\sqrt{J_{min}J_{max}}
\end{eqnarray}
This expression has a particularly transparent meaning for the unidirectional
``striped-like'' variation of $U(\br)$ that we treated explicitly when solving
the mean-field equations, above.  There, $J_{max}$ corresponds to the stiffness
along the stripes and $J_{min}$ -- perpendicular to the stripes.  The
corresponding anisotropic XY model directly leads to the result
Eq.~(\ref{eq:T_KT}).  In this case, we find
\begin{eqnarray}
T_{KT} \sim \frac {T_f}{T_{KT}^2} T_c^{MF} \Delta_{min}(T_{KT}) .\nonumber
\end{eqnarray}
Together with Eq.~(\ref{eq:min2}) for $\Delta_{min}(T_{KT}) \sim \Delta(L,
T_{KT})$, this equation implicitly defines $T_{KT}$.  With logarithmic accuracy
we find that $T_{KT}\sim \min(T_c^{MF}, v_fQ/A)$.

In any case, baring certain artificial geometries, it  is clear that for a long
wave length modulation, $\xi Q \ll 1$, the Kosterlitz-Thouless temperature
$T_{KT}$ is exponentially lower than the MF transition temperature; on the
other hand, for modulations with $\xi Q \sim 1$, the phase fluctuation region
is very narrow and $T_{KT}\approx T_c^{MF}$.  In this regime, the mean-field
superconducting temperature is still exponentially enhanced relative to its
value in the uniform state with the same average paring interaction strength,
${\bar U}$.  For even faster modulation, $\xi Q \gg  1$, the MF transition
temperature drops since the pairings interaction modulation averages out on the
length scale of $\xi$. This trend is presented qualitatively in
Fig.~\ref{fig:Tc}.

For a ``dirty'' superconductor with a  mean free path shorter than the clean
coherence length, $\ell = v_F \tau < \xi$, the effect of phase fluctuations can
be estimated in the same way as in the clean limit, with minor changes
(involving prefactors) but with the coherence length redefined as $\xi_d =
\sqrt{\xi \ell}$.

{\em Summary.} We studied the effect of nanoscale inhomogeneity on the
superconducting transition temperature, $T_c$.  We considered two possible
kinds of inhomogeneity:  the modulations of the paring strength and of the
background potential. In the weak coupling BCS regime, we find that
inhomogeneity which is incommensurate with the Fermi surface nesting vectors
$enhances$ $T_c$ relative to its value for the uniform zero center-of-mass
momentum pairing. For a fixed modulation depth we find that the highest $T_c$
is reached when the modulation wavelength is of the order of the
superconducting coherence length. For shorter wavelengths, the superconductor
cannot take advantage of the locally favorable conditions, while for the longer
wavelengths, the global superconductivity is suppressed due to the phase
fluctuations on the weak links, where the amplitude of the order parameter is
significantly reduced.  Although explicitly derived for $s$-wave
superconductors, similar results will also apply to unconventional
superconductors in the presence of smooth (on the $1/k_f$ length scale)
inhomogeneities.  Clearly oversimplified, the presented picture bears
resemblance to the high-temperature superconducting cuprates, where
considerable experimental evidence\cite{emery} indicates that the maximum $T_c$
occurs at a crossover between a regime where $T_c$ is controlled by the pairing
scale and where it is a phase ordering transition.

We thank G. Blumberg, S. Chakravarty, G. Ortiz, M. Stepanov, and L. Gruvitz for
useful discussions. We also acknowledge hospitality of Aspen Center for Physics
where part of this work was completed. IM and DP were supported by the US DoE.
SAK was supported by NSF grant No. DMR 0421960. 


\vspace{-4 mm}

\begin{thebibliography}{9}
\vspace{-3 mm}
\bibitem{davis}  J. E. Hoffman {\em et al.}, Science {\bf 295}, 466 (2002); T.
Hanaguri {\em et al.}, Nature {\bf 430}, 1001 (2004).
\bibitem{kapit} C. Howald {\em et al.}, Phys. Rev. B {\bf 67}, 014533(2003).
\bibitem{yazd} Vershinin {\em et al.}, Science {\bf 303}, 1995 (2004).
\bibitem{kiv1} S. A. Kivelson and V. J. Emery, in {\it Strongly Correlated Electronic
Materials: The Los Alamos Symposium 1993} ed. by K. S. Bedell, Z.
Wang, B. E. Meltzer, A.V. Balatsky, and E. Abrahams
(Addison-Wesley, Redding, 1994).
\bibitem{kiv2} E. Arrigoni, E. Fradkin, and S. A. Kivelson,  Phys. Rev. B {\bf 69},
214519 (2004).
\bibitem{erica} E. W. Carlson, V. J. Emery, S. A. Kivelson, and D.
Orgad, in {\it The Physics of Conventional and Unconventional
Superconductors}, edited by K. H. Bennemann and J. B. Ketterson
(Springer-Verlag, Berlin, 2002).
\bibitem{machida} K. Machida, T. Koyama, and T. Matsubara, Phys. Rev. B {\bf
23}, 99 (1981).
\bibitem{sach} M. Vojta and S. Sachdev, Phys. Rev. Lett. {\bf 83}, 3916 (1999).
\bibitem{ivar} I. Martin, G. Ortiz, A. V. Balatsky, and A. R. Bishop, Int. J. of Mod. Phys., {\bf 14}, 3567
(2000).
\bibitem{ddw} S. Chakravarty, R. B. Laughlin, D. K. Morr, C. Nayak, Phys. Rev. B {\bf 63},
094503 (2001).
\bibitem{agd} A. A. Abrikosov, L. P. Gorkov, and I. E. Dzyaloshinski, {\em Methods of Quantum Field Theory in Statistical
Physics} (Dover, New York, 1975).
\bibitem{Bellman} R. Bellman, {\em Introduction to Matrix Analysis} (Soc for Industrial \& Applied Math,
1997).
\bibitem{MW} N. D. Mermin and H. Wagner, Phys. Rev. Lett. {\bf 22}, 1133
(1966).
\bibitem{KT} J. M. Kosterlitz and D. J. Thouless, J. Phys. C {\bf 6}, 1181
(1973).
\bibitem{emery}  V.J.Emery and S.A.Kivelson, {\it Nature} {\bf 374} 434 (1995).

\end{thebibliography}
\end{document}